\documentstyle[12pt]{article}
  
\def\half{{\textstyle {1 \over 2}}}

\def\rchi{{\raise 2pt \hbox {$\chi$}}}
\def\rga{{\raise 2pt \hbox {$\gamma$}}}
\def\rg{{\raise 2 pt \hbox {$g$}}}

\def\susy{supersymmetry}
 
\def\<{\left\langle}
\def\>{\right\rangle}

\def\pt{\partial}
\def\eps{\epsilon}

\def\ol{\overline}

\def\lam{\lambda}

\def\sig{\sigma}

\def\ti{\tilde}

\begin{document}

\title{{\Large\bf  Quantization of a 
\\  Friedmann-Robertson-Walker  Model \\ with Gauge Fields \\ 
in N=1 Supergravity }}
\author{{\large\sf P.V. Moniz}\thanks{\sf e-mail: prlvm10@amtp.cam.ac.uk; PMONIZ@Delphi.com}\\
Department of Applied Mathematics and Theoretical Physics \\
University of Cambridge \\ Silver Street, Cambridge, CB3 9EW\\ 
United Kingdom}
\date{ DAMTP R95/36}
\maketitle

\begin{abstract}

The purpose of this paper is to investigate a specific FRW 
model derived from the theory of N=1 supergravity with gauged supermatter. 
The supermatter content is restricted to a vector supermultiplet. This objective 
is particularly worthwhile. In fact, it was pointed in ref. \cite{24} 
({\em Class. Quantum Grav. {\bf 12} {\rm (} {\rm 1995} {\rm )} 
{\rm 1343}}) 
that $\Psi = 0$ was the only allowed quantum state 
for N=1 supergravity with {\em generic} gauged supermatter 
subject to suitable    FRW ans\"atze. 

The ans\"atze employed here for the  physical variables 
was presented in 
ref. \cite{24}. 
The corresponding  Lorentz and supersymmetry quantum constraints are then  derived. 
Non-trivial solutions are subsquently found. A no-boundary solution is 
identified while another state may be interpreted as a wormhole solution. 
In addition, the usefulness and limitations of the  ans\"atze 
are addressed. The implications of the ans\"atze with 
respect to  the allowed 
quantum states are also discussed with a view on ref. \cite{24}.

  \end{abstract}

\section{Introduction}

\indent


Research in supersymmetric quantum
 cosmology  using 
canonical methods started about 20 years or so 
\cite{1}-\cite{3}. Since then, many other papers have appeared in 
the literature   
\cite{5}-\cite{30}. A  review on the subject 
of canonical quantum N=1,2 supergravity 
can be found in 
ref. \cite{31}.

 An important feature
of 
  N=1 supergravity is that it constitutes a ``square-root''  
  of gravity
\cite{1}. In fact, the Lorentz and supersymmetry 
constraints of the theory induce   a set of coupled 
first-order  differential equations   
which the quantum states ought to satisfy. 
Such physical states will subsequently  
 satisfy the Hamiltonian constraints as a consequence 
of 
the
 supersymmetry algebra \cite{1}-\cite{3}.  
Supersymmetry may also play an important role when dealing with (ultraviolet) 
divergences in quantum cosmology and gravity \cite{4}
and removing Planckian masses 
induced by wormholes \cite{5,6}. Hence,   canonical quantization 
methods may  provide new and interesting 
insights as far as quantum supergravity theories are  concerned.

Important results 
for Bianchi class-A models were recently achieved 
within pure N=1 supergravity. 
On the one hand, the Hartle-Hawking (no-boundary) \cite{32} and  
wormhole (Hawking-Page)  \cite{33} states 
 were 
found  in the {\it same} spectrum of solutions  \cite{14,15}. This result improved previous attempts \cite{7}-\cite{13} 
where {\it only one} 
 of these   states could be found, depending on the homogeneity conditions 
imposed on the gravitino \cite{9}. 
The reason was an overly restricted ansatz for 
the wave function of the universe. 
More precisely, gravitational degrees of freedom have not 
been properly taken into account in the   expansion of $\Psi$  
in Lorentz invariant fermionic sectors.  
Another undesirable consequence of the ansatz used in \cite{7}-\cite{13} 
occured from the inclusion of a cosmological constant \cite{16}-\cite{18}:
   Bianchi class-A models were 
found to have  {\it no} 
physical states   but the trivial one, $\Psi =0$. However, 
 an extension of the method 
in \cite{14,15} 
 did led in ref. \cite{19} 
to solutions of the form of  exponentials of the Chern-Simons functional.

The introduction of supermatter 
\cite{20,21} in supersymmetric minisuperspaces 
provided  
challenging  results. A scalar supermultiplet, 
constituted by  complex scalar fields, $\phi,  \bar \phi$ and their 
spin$-{1 \over 2}$ partners, $\chi_A, \bar \chi_{A'}$ was considered in ref. 
\cite{5,11,21,22}, \cite{25}-\cite{28}. 
A vector supermultiplet, formed by a gauge vector field $A^{(a)}_\mu$ and its 
supersymmetric partner,  was added   in ref. \cite{23,24}.
A FRW model was considered in ref. \cite{5,11,21,22,25,26} 
 while
 a Bianchi IX model was dealt with in ref. \cite{27,28}. 
Bianchi class-A models with Maxwell fields within N=2 supergravity were 
considered in ref. \cite{29,30}. 

\vspace{0.2cm}

Finding  and identifying  physical states 
in  minisuperspaces obtained from 
supergravity theories with supermatter constitutes 
an  important assignement:

\begin{itemize}

\vspace{0.2cm}

\item A wormhole 
  state was obtained in ref. \cite{5},
 but {\it not} 
 in ref. \cite{22}. We emphasize that the more general theory 
of N=1 supergravity with 
gauged supermatter \cite{34} was  employed in ref. \cite{22}. The reason for the 
discrepancy between \cite{5,22} 
 was analysed  in ref. \cite{25,26}  and related to the 
type of Lagrange multipliers and fermionic derivative ordering that 
were used. 

\item As far as a Hartle-Hawking state is concerned, some 
 solutions present in the 
literature 
bear some of the properties corresponding to the no-boundary proposal
\cite{32}. 
Unfortunately,   the supersymmetry 
constraints were  {\it not} 
 sufficient in determining the dependence of $\Psi$ with respect to 
the scalar field (cf. ref. \cite{25,26} for more details). 

\item The results found within the 
 more general  matter content used in ref. \cite{23,24}   were 
 disapointing:   the only allowed  physical state was $\Psi = 0$. 
 
\end{itemize}

It is of interest to address some of these problems. 
 On the one hand,  the apparent absence of 
wormhole solutions \cite{25,26} 
and the difficulty to  obtain  an adequate  Hartle-Hawking solution
when scalar supermultiplets are considered. On the other hand,
 why non-trivial physical states are not permited 
when all possible matter fields are present \cite{23,24}.

\vspace{0.3cm}

The main purpose of this paper 
 is precisely   to  
investigate a particular 
FRW model obtained from 
N=1 supergravity with supermatter restricted 
to a  vector supermultiplet. 
As a consequence, we hope to shed some light on the  
the parodoxical situation found in ref. \cite{23,24},  
in spite of all the degrees of freedom thereby present.
 
  In section 2 we will address the ans\"atze 
for the field variables. We summarize in subsection 2.1 the basic 
properties of the ans\"atze employed in the dimensional reduction 
of pure N=1 supergravity to a FRW model 
(see ref.  \cite{10,11}).   In subsection 2.2 we will consider 
our  FRW case model. 
We put all scalar fields and corresponding supersymmetric 
partners equal to zero. The ans\"atze 
employed here (and in ref. \cite{23,24}) 
for the spin-1 field 
and corresponding fermionic partners 
are then   described   in some detail (see  ref. \cite{35}-\cite{39} 
as well). 
The ans\"atze introduced in ref. 
\cite{10,11} for the case of N=1 pure supergravity 
 are preserved under a combination of 
supersymmetry, local coordinates and Lorentz transformations. 
The   preservation of our    ans\"atze 
under combined gauge, 
supersymmetry, local coordinates and Lorentz transformations
can  be achieved only after {\em further}  restrictions are imposed. 
It should be emphasized that a discussion on {\em how} gravitational, 
gravitino and supermatter fields can be accommodated within a 
$k=1$ FRW geometry and supersymmetry was {\em not} performed in 
ref. \cite{23,24}. Thus, the content of  subsection 2.2 and following sections 
will surely contribute to our knowledge on supersymmetric minisuperspaces 
and associated features.

In section 3 we will derive the corresponding  
Lorentz and supersymmetry constraints. 
The results subsequently found 
will prove to be physically  
interesting.  
In contrast with ref. \cite{23,24},   {\it non-trivial} solutions in different 
fermionic sectors are   obtained. We identify a {\it component} of 
the Hartle-Hawking (no-boundary) 
solution 
 \cite{32,35}. Another 
solution 
  could be interpreted as a quantum wormhole state 
\cite{33} (see  ref.  \cite{35a}).
We stress  that the Hartle-Hawking solution found here 
is 
part of the set of   solutions 
shown in ref. \cite{35}, where a non-supersymmetric 
 FRW 
minisuperspace with Yang-Mills fields was considered. Being N=1 
supergravity a square-root of gravity, this result  is 
particularly supportive.  
It further points that our approach and ans\"atze (see ref. 
\cite{23,24})  
may    be of some utility. 
Finally, our discussions and 
conclusions   close this paper in section 4.

\section{Ans\"atze for the field variables}

\indent

The action for our model 
 is obtained from the more general theory of 
N=1 supergravity with gauged supermatter 
\cite{34}. We  put all  scalar fields and corresponding 
supersymmetric partners equal to zero 
(it should be noticed that  Yang-Mills fields coupled to N=1 supergravity
can be found in ref. \cite{40}). 
The corresponding  field variables will be constituted by a tetrad, 
$e^{AA'}_\mu$ (in two spinor component notation  
- cf.,  e.g.,  ref. \cite{3}),  gravitino fields, 
$\psi^A_\mu, \bar \psi^{A'}_\nu$,  (where a bar denotes Hermitian conjugation), 
a gauge spin$-1$ field, $A^{(a)}_\mu$, (where $(a)$ is a gauge group index) and the the 
corresponding spin-${1 \over 2}$  partners, $\lambda^{(a)}_A, \bar \lambda^{(a)}_{A'}$.

The restriction of this theory to a closed FRW model requires the 
introduction of  specific ans\"atze for the fields mentioned above.
This is  discussed  
in the following subsections.
We will start by reviewing how the tetrad and the gravitino fields can be chosen 
in pure N=1 supergravity \cite{10,11}. 

\subsection{FRW model from pure N=1 supergravity}

\indent

We choose
 the geometry 
to be that of a $ k = + 1 $ Friedmann model with $ S^3 $ spatial sections, 
which are the spatial
 orbits of $G=SO(4)$ -- the (isometry) group of homogeneity and 
isotropy. 
 The ansatz for the tetrad can then be written as 
\begin{equation}
 e_{a\mu} = \left(\begin{array}{cc}
N (\tau) &0  \\
0 & a (\tau) E_{\hat a i} \end{array}\right)~,
~e^{a \mu} = \left( \begin{array}{cc}
N (\tau)^{-1} & 0 \\
0 & a (\tau)^{-1} E^{\hat a i} \end{array} \right)~,
\label{eq:1}
\end{equation}
where $ \hat a $ and $ i $ run from 1 to 3.  
$ E_{\hat a i} $ is a basis of left-invariant 1-forms on the unit $ S^3 $
with volume $ \sig^2 = 2 \pi^2 $. The spatial tetrad $ e^{AA'}_{~~~~i} $
satisfies the relation
$ \pt_i e^{AA'}_{~~~~j} - \pt_j e^{AA'}_{~~~~i} = 2 a^2 e_{ijk} e^{AA'k}$
as a consequence of the group structure of SO(3), the isotropy 
(sub)group.

This ansatz reduces the number of degrees of freedom provided by $ e_{AA'
\mu} $. If supersymmetry invariance is to be retained, then we need an
Ansatz for $ \psi^A_{~~\mu} $ and $ \bar\psi^{A'}_{~~\mu} $ which reduces the 
number of fermionic degrees of freedom as well   \cite{10,11}. The  
Lagrange multipliers 
$
\psi^A_{~~0} $ and $ \bar\psi^{A'}_{~~0} $ are taken 
to be functions of time only. 
The Ansatz for the gravitino field further includes 
\begin{equation} 
\psi^A_{~~i} = e^{AA'}_{~~~~i} \bar\psi_{A'}~, ~
\bar\psi^{A'}_{~~i} = e^{AA'}_{~~~~i} \psi_A~, \label{eq:2}
\end{equation}
where we introduce the new spinors $ \psi_A $ and $ \bar\psi_{A'} $ which
are functions of time only. 
This means we truncate the general decomposition
$ \psi^A_{~~B B'} = e_{B B'}^{~~~~i} \psi^A_{~~i} $, 
\begin{equation}
 \psi_{A B B'} = - 2 n^C_{~~B'} \rga_{A B  C} + {2 \over 3} \left( \beta_A
n_{B B'} + \beta_B n_{A B'} \right) - 
2 \varepsilon_{A B} n^C_{~~B'} \beta_C~, \label{eq:3}
\end{equation}
where $ \rga_{A B C} = \rga_{(A B C)} $,  
at spin$-\half$ mode level. I.e.,  
$\beta^A = {3 \over 4} n^{AA'} \ol\psi_{A'} \sim \ol\psi^A$.
However,  it is  important to stress that 
auxialiary fields were further  
required in \cite{10,11} to balance the number of 
degrees of freedom in ans\"atze (\ref{eq:1}), 
(\ref{eq:2}).

Ans\"atze  (\ref{eq:1}), 
(\ref{eq:2}) are preserved by a combination of local coordinate 
 - $\delta_{(lc)} $ - , Lorentz  - $\delta_{(L)} $ -  and supersymmetry  - $\delta_{(s)}
 $ -  
transformations. In the 
4-dimensional theory, these transformations are given by  
\begin{eqnarray}
\delta_{(lc)} e^{AA'}_{~~~~\mu} & = &\xi^\nu \partial_\nu e^{AA'}_{~~~~\mu} + 
e^{AA'}_{~~~~\nu} \partial_\mu \xi^\nu~, 
\label{eq:4}\\
\delta_{(L)} 
 e^{AA'}_{~~~~\mu} & = &N^A_{~~B} e^{BA'}_{~~~~\mu} + \bar N^{A'}_{~~B'}
e^{AB'}_{~~~~\mu}~, \label{eq:5} \\
\delta_{(s)} e^{AA'}_{~~~~\mu} &=& - i \left( \epsilon^A \bar \psi^{A'}_{~~\mu} .+ 
\bar \epsilon^{A'} \psi^A_{~~\mu} \right)~, \label{eq:6}  \\
\delta_{(lc)} \psi^A_{~~\mu} &=& \xi^\nu \partial_\nu \psi^A_{~~\mu} + \psi^A_{~~\nu} \partial_\mu 
\xi^\nu~, \label{eq:7} \\
\delta_{(L)} \psi^A_{~~\mu} &=&N^A_{~~B} \psi^B_{~~\mu}~, 
\label{eq:8} \\
\delta_{(s)} \psi^A_\mu &= & 2D_\mu \epsilon^A = 2 \pt_\mu \epsilon^A + \omega^A_{~B\mu} 
\epsilon^B ~,\label{eq:9} 
\end{eqnarray}
with the Hermitian conjugates of (\ref{eq:7}), (\ref{eq:8}), (\ref{eq:9}) and 
where $\xi^\mu, N^{AB}, \epsilon^A$ are time-dependent vectors or 
spinors parametrizing, respectively, the above mentioned transformations.

In order for the ansatz (\ref{eq:1}) to be preserved under a combination 
of transformations (\ref{eq:4}), (\ref{eq:5}), (\ref{eq:6}) it ought to satisfy 
a relation as 
\begin{equation}
\delta e^{AA'}_i =  P_1\left[e^{AA'}_\mu, \psi^A_\mu\right] 
e^{AA'}_i,
\label{eq:10}
\end{equation}
where 
$ P_1$ is an  expression 
(spatially independent and possibly complex) 
where all spatial and spinorial indices have been contracted. 
Using 
expressions
(\ref{eq:4}), (\ref{eq:5}), (\ref{eq:6}) with ansatz (\ref{eq:2}), defining
\begin{equation}
\xi^i = - \xi^{AA'}e_{AA'}^i ~,~ \xi^{AA'} = i\xi^{AB} n_B^{A'} ~,~
\bar\xi^{A'B'} = 2 \xi^{AB} n_A^{B'} n_B^{B'}~,
\label{eq:11}
\end{equation} 
and putting $\xi^0 = 0$ \cite{11}, we get 
\begin{eqnarray}
\delta e^{AA'}_{~~~~i} &= &   - N^{AB} e_{iB}^{~~A'} - \bar N^{A'B'}
e^B_{~B'i} \nonumber \\
& + &  a^{-1} \xi^{AC} e_{iC}^{~~A'} + 
a^{-1} \tilde \xi^{A'C'} e^A_{~C'i} \nonumber \\
&+&  i   \left(
\epsilon^{A} \psi^{B}e_{iB}^{~~A'} +  \bar\epsilon^{A'} \bar
\psi^{B'} e^A_{~B'i}\right)~. 
\label{eq:12}
\end{eqnarray}
Using now the spinorial relation
\begin{equation}
\psi^{[A} \chi^{B]} = \half \epsilon^{AB} \psi_C\chi^C~,
\label{eq:13}
\end{equation}
we derive that 
\begin{equation}
\epsilon^A\psi^B e_{iB}^{~A'} = \varepsilon^{(A}\psi^{B)} e_{iB}^{~A'} 
+ \half \epsilon_C \psi^C e^{~AA'}_i~,
\label{eq:14}
\end{equation}
from which it follows that \cite{10,11}
\begin{eqnarray}
\delta e^{AA'}_{~~~~i} &= &  \left( - N^{AB} + a^{-1} \xi^{AB}
+ i  \epsilon^{(A} \psi^{B)} \right ) e_B^{~~A'}{}_i  \nonumber \\
~&+&  \left( - \bar  N^{A'B'} + a^{-1} \bar  \xi^{A'B'} + i
\bar \epsilon^{(A'} \bar
\psi^{B')} \right) e^A_{~~B'i} \nonumber \\
~&+&  {i  \over 2} \left( \epsilon_C \psi^C + \bar \epsilon_{C'} \bar \psi^{C'} \right) 
e^{AA'}_{~~~~i}~.
\label{eq:15}
\end{eqnarray}
The relation (\ref{eq:10}) holds 
provided that the relations
\begin{equation}
N^{AB} - a^{-1} \xi^{AB} - i  \epsilon^{(A} \psi^{B)} = 0~, 
\bar N^{A'B'} - a^{-1} \bar \xi^{A'B'} - i \bar
\epsilon^{(A'} \bar \psi^{B')} = 0~, \label{eq:16}
\end{equation}
between the generators of Lorentz, coordinate and supersymmetry 
transformations are satisfied.
Hence, we will achieve $\delta e^{AA'}_{~~i} = 
C(t) \delta e^{AA'}_{~~i}$ with  $C(t) = 
{i  \over 2} \left( \epsilon_C \psi^C + \bar \epsilon_{C'} \bar \psi^{C'} \right)$
 and the 
ansatz (\ref{eq:1}) remains unchanged (see ref. \cite{11}). Notice that 
any Grassman-algebra-valued field can be decomposed into a ``body'' or 
component along unity (which takes values in the 
field of real or complex numbers) and a ``soul'' which is nilpotent 
(see ref. \cite{nilp}). 
The variation (\ref{eq:15}) with conditions (\ref{eq:16}) imply that 
$\delta e^{AA'}_{~~i}$ exists entirely in the nilpotent (``soul'') part.

Let us now address ansatz (\ref{eq:2}). 
From   eq.  (\ref{eq:1}), (\ref{eq:2}) we 
get (cf. \cite{11})
\begin{eqnarray}
2D_i \epsilon^A &= &   \left[ 2 \left( {\dot a \over a N} + {i \over a} \right) 
- {i \over
2 N} \left( \psi_F \psi^F_{~~0} + \bar
 \psi_{F'0} \bar \psi^{F'} \right) \right] n_{BA'} 
e^{AA'}_{~~~~i} \eps^B \nonumber \\
& + & \frac{i}{2} e^{(A\mid E'\mid}_{~~~~i}
\varepsilon^{B)E}\psi_E\bar\psi_{E'}\epsilon_B~.
\label{eq:17}
\end{eqnarray}
Using then eq. (\ref{eq:11}), (\ref{eq:13}), (\ref{eq:14}), (\ref{eq:16}) and 
(\ref{eq:17}) it follows that 
\begin{eqnarray}
\delta \psi^A_{~~i} & = &{3i \over 4} \epsilon^A \psi^B \bar \psi^{B'} e_{BB'i} + 
a^{-1} \bar \xi^{A'B'} e^A_{~~B'i} \bar \psi_{A'} \nonumber \\
~&+ & \left[ {2} \left( {\dot a \over a N} + {i \over a} \right) - {i  \over
2 N} \left( \psi_F \psi^F_{~~0} + \bar \psi_{F'0} \bar \psi^{F'} \right) \right] n_{BA'} 
e^{AA'}_{~~~~i} \eps^B~. 
\label{eq:18}
\end{eqnarray}
Notice the factor $\frac{3}{4}$ as different from \cite{11}. 
Hence,   to recover 
for eq. (\ref{eq:18}) a relation like   
\begin{equation}
\delta \psi^A_i =  
P_2 \left[e^{AA'}_\mu, \psi^A_\mu\right]
e^{AA'}_i\bar \psi_{A'}, 
\label{eq:20}
\end{equation} and its Hermitian conjugate (i.e., 
in order for ansatz (\ref{eq:2}) to be unchanged)  
the following conditions ought to be used.

First, we  require $\xi^{A'B'} = \xi^{AB} = 0$ (see ref. \cite{11}). 
 In addition, we require that 
\begin{equation}
\left[ {2  } \left( {\dot a \over a N} + {i \over a} \right) - {i  \over
2 N} \left( \psi_F \psi^F_{~~0} + \bar
 \psi_{F'0} \bar \psi^{F'} \right) \right] n_{BA'} 
 \eps^B~ \sim  P_2 \left[e^{AA'}_\mu, \psi^A_\mu\right] \bar\psi_{A'}.
\label{eq:21}
\end{equation}
This means that the variation  $\delta \psi^A_i = D(t) \psi^A_i$  
with $D(t) = \left[ {2  } \left( {\dot a \over a N} + {i \over a} \right) - {i  \over
2 N} \left( \psi_F \psi^F_{~~0} + \bar
 \psi_{F'0} \bar \psi^{F'} \right) \right]$
  will have a component along unity 
(``body'' of Grassman algebra) and another which is nilpotent 
(the ``soul'': $ \psi_F \psi^F_{~~0} + \bar
 \psi_{F'0} \bar \psi^{F'} $).
Finally, 
the additional 
constraint
$ \psi^B \bar\psi^{B'} e_{BB'i} = 0$ follows.
This amounts to $ \psi^B \bar\psi^{B'} \propto n^{BB'} $, 
and so we can write  
\begin{equation}
J_{AB} = \psi_{(A} \bar\psi^{B'} n_{B)B'} = 0~.
\label{eq:22}
\end{equation}
 The 
constraint $ J_{AB} = 0 $ has a natural interpretation as the reduced form of 
the Lorentz rotation constraint arising in the full theory. 
By requiring that the constraint $ J_{AB} = 0 $ be preserved under the same 
combination of transformations as used above, one finds equations
 which 
are satisfied provided the supersymmetry constraints $ S_A = 0,~\bar S_{A'} =
0 $  hold. 
By further requiring that the supersymmetry 
constraints be preserved, one finds additionally that the Hamiltonian 
constraint $ {\cal H} = 0 $  should hold. 
The invariance of $\bar\psi^{A'}_i$ under 
transformations (\ref{eq:7}), (\ref{eq:8}), (\ref{eq:9}) leads to the 
Hermitian conjugates of the above expressions.

In the next section we will discuss in detail our ans\"atze 
(see ref. \cite{23,24}) for the physical variables of a $k=1$ FRW 
model within N=1 supergravity with a vector supermultiplet. 
Namely, our objective is to access {\it how} these  
ans\"atze can be accommodated with both the FRW geometry and 
supersymmetry. In doing so we aim to provide relevant information that 
is absent in ref. \cite{23,24} and thus to 
improve our understanding of supersymmetric minisuperspaces 
and their current problems.  

\subsection{FRW model with gauge fields from N=1 supergravity 
with supermatter}

\indent

Within the more general  theory of N=1 supergravity with gauged supermatter, 
our field variables (together with scalar fields $\phi^I, \phi^{J*}$ 
and fermionic partners $\chi_A^J, \bar\chi_{A'}^{J*}$)
are transformed as follows \cite{34}. Under supersymmetry transformations - $\delta_{(s)}$ -
we have 
\begin{eqnarray} 
\delta_{(s)} e^{AA'}_{~~~~\mu} &=& - i \left( \epsilon^A \bar \psi^{A'}_{~~\mu} .+ 
\bar \epsilon^{A'} \psi^A_{~~\mu} \right)~, \label{eq:23} \\
\delta_{(s)} \psi^A_{~~\mu} &=  &2  \hat{D}_\mu \epsilon^A
 -{i \over 2} C_{\mu\nu}^{~~AB} \epsilon_B g_{IJ^*}\chi^{IC} e^{\nu}_{CC'}
\bar\chi^{J^{*}C'} \nonumber  \\
& + & {i\over 2} \left(g_{\mu\nu} \varepsilon^{AB} + 
C_{\mu\nu}^{~~AB}\right)\epsilon_B \lambda^{(a)C} e^\nu_{CC'} \bar\lambda^{(a)C'} 
\nonumber \\
&- &{1 \over 4} \left( {\partial K \over \partial \phi^J} \delta_{(s)}\phi^J 
- {\partial K \over \partial \bar\phi^{J^{*}}} \delta_{(s)}\bar\phi^{J^{*}}\right)\psi^A_\mu 
+ ie^{K/2} P e_{\mu}^{ AB'} \bar\epsilon_{B'}~,  \label{eq:24} \\
\delta_{(s)} A_\mu^{(a)} &= & i \left(\epsilon^A e_{\mu AA'} \bar\lambda^{(a)A'} - 
\lambda^{(a)A} e_{\mu AA'} \bar\epsilon^{A'}\right) , \label{eq:25}\\
\delta_{(s)}  \lambda^{(a)A}   &= &- {1 \over 4} \left( {\partial K \over \partial \phi^J} \delta_{(s)}\phi^J 
- {\partial K \over \partial \bar\phi^{J^{*}}} \delta_{(s)}\bar\phi^{J^{*}}\right)\lambda^{(a)A} -
 iD^{(a)} \epsilon_A + \hat{F}^{(a)}_{\mu\nu} C^{\mu\nu~~B}_{~~~A} \epsilon_B
 ~,  \label{eq:26} \\
\delta_{(s)}\phi^I &=& \sqrt{2} \epsilon_A \chi^{IA},~ \label{eq:27}\\
\delta_{(s)} \chi^{I}_A &= &i\sqrt{2} e^\mu_{AA'} \bar\epsilon^{A'} 
\left(\tilde D_\mu \phi^I 
-\frac{1}{2} \sqrt{2} \psi_{\mu C}\chi^{IC}\right) - \Gamma^I_{JK} \delta_{(s)}\phi^J \chi^K_A \nonumber \\
&+ &{1 \over 4} \left( {\partial K \over \partial \phi^J} \delta_{(s)}\phi^J 
- {\partial K \over \partial \bar\phi^{J^{*}}} \delta_{(s)}\bar\phi^{J^{*}}\right)\chi^I_A 
-\sqrt{2} e^{K/2} g^{IJ^{*}}D_{J^{*}}P^*\epsilon_A
~, \label{eq:28} \end{eqnarray}
and their Hermitian conjugates. Here 
 $\varepsilon^{AB}$ is the alternating spinor, 
 $ \epsilon^A $ and $ \bar \epsilon^{A'} $ are odd (anticommuting) fields, 
$C^{\mu\nu~~B}_{~~A} = {1 \over 4} 
(e^\mu_{AA'}e^{\nu BA'} - e_{AA'}^\nu e^{\mu BA'})$, 
$\hat{F}_{\mu\nu}^{~(a)} = F^{~(a)}_{\mu\nu} $ $- 
i\left(\psi^A_{[\mu}e_{\nu]AA'} \bar\lambda^{(a)A'}
\right.$ $\left. - \bar\psi_{A'[\mu}e_{\nu]}^{AA'}\lambda^{(a)}_A\right)$. 
$\Gamma^I_{JK}$ is a Christoffel symbol derived from 
K\"ahler metric $g_{IJ^{*}} = 
{\partial^2 K \over {\partial \phi^I \partial \bar\phi^{J*}}}$, 
$K$ is the K\"ahler potential, $P$ is a complex scalar-field dependent 
potential energy term, $D^{(a)}$ is a Killing potential and $X^{I(a)}$ 
is a Killing vector on the K\"ahler manifold. In addition, 
$D_I = {\partial \over \partial \phi^{I}} + {\partial K \over \partial \phi^{I}}$, 
$\hat{D}_\mu = \partial_\mu + \omega_\mu + 
{1 \over 4} \left( {\partial K \over \partial \phi^J} \tilde D_\mu\phi^J 
- {\partial K \over \partial \bar\phi^{J^{*}}} \tilde D_\mu\bar\phi^{J^{*}}\right) 
+ {i\over 2} A^{(a)}_\mu Im F^{(a)}$, 
$\tilde D_\mu = \partial_\mu - A^{(a)}_\mu X^{I(a)}$,  
$F^{(a)} = X^{I(a)}{\partial K \over \partial \phi^I} + i D^{(a)}$  and
$I$ denotes K\"ahler indices.

For Lorentz transformations - $\delta_{(L)}$ - it follows that 
\begin{eqnarray}
\delta_{(L)} 
 e^{AA'}_{~~~~\mu} & = &N^A_{~~B} e^{BA'}_{~~~~\mu} + \bar N^{A'}_{~~B'}
e^{AB'}_{~~~~\mu}~, \label{eq:28a} \\
\delta_{(L)} \psi^A_{~~\mu} &=&N^A_{~~B} \psi^B_{~~\mu}~, 
\label{eq:29} \\
\delta_{(L)} \phi^I &=& 0,~~ \delta_{(L)} \chi^I_A = N_{AB}\chi^{IB},~ 
\label{eq:30}\\
\delta_{(L)} A^{(a)}_\mu &= &N_\mu^{~\nu} A^{(a)}_\nu, ~
\delta_{(L)} \lambda^{(a)}_A = N_{AB} \lambda^{(a)B} 
\label{eq:31}
 \end{eqnarray} with their 
Hermitian 
conjugates,  where $ N^{AB} = N^{(AB)} $ and $  N^{\mu\nu} = N^{[\mu\nu]} $.

Under local coordinate transformations - $\delta_{(lc)}$ - we have 
\begin{eqnarray} 
\delta_{(lc)} e^{AA'}_{~~~~\mu} & = &\xi^\nu \partial_\nu e^{AA'}_{~~~~\mu} + 
e^{AA'}_{~~~~\nu} \partial_\mu \xi^\nu~, 
\label{eq:32}\\
\delta_{(lc)} \psi^A_{~~\mu} &=& \xi^\nu \partial_\nu \psi^A_{~~\mu} + \psi^A_{~~\nu} \partial_\mu 
\xi^\nu~, \label{eq:33} \\
\delta_{(lc)} A^{(a)}_\mu &= &\xi^\nu \partial_\nu A^{(a)}_\mu 
+ A^{(a)}_\nu \partial_\mu 
\xi^\nu~, \label{eq:34}\\
\delta_{(lc)} \lambda^{(a)}_A &= &\xi^\nu \partial_\nu \lambda^{(a)}_A~,
\label{eq:35} \\
\delta_{(lc)} \phi^I&= &\xi^\nu \partial_\nu \phi^I, ~ 
\delta_{(lc)} \chi^I_A = \xi^\nu \partial_\nu \chi^I_A, \label{eq:36}
  \end{eqnarray} considering the Hermitians conjugates as well.

Finally,  for gauge transformations - $ \delta_{({\rm g})} $ - we get 
\begin{eqnarray}
\delta_{({\rm g})} \psi^A_\mu &=& -{i\over 2} \zeta^{(a)} Im F^{(a)} 
\psi^A_\mu,~~\label{eq:37} \\
\delta_{({\rm g})} \chi^I_A &= &\zeta^{(a)} {\partial X^{I(a)} \over \phi^{J}} 
\chi^J_A + {i\over 2} \zeta^{(a)} Im F^{(a)} \chi^I_A,~
\delta_{({\rm g})} \phi^I = \zeta^{(a)} X^{I(a)},~
\label{eq:38}\\
\delta_{({\rm g})} A^{(a)}_\mu &=& \partial_\mu \zeta^{(a)} + k^{abc} 
\zeta^{(b)} A_\mu^{(c)},~\label{39} \\
\delta_{({\rm g})} \lambda^{(a)}_A &= &
k^{abc} \zeta^{(b)} \lambda ^{(c)}_A - {i\over 2}\zeta^{(b)} Im F^{(b)} 
\lambda^{(b)}_A, \label{eq:40}
  \end{eqnarray}
and Hermitian conjugates.

\vspace{0.4cm}

Let us then consider 
the case\footnote{{\rm 
For the case of a  gauge group SO(3) $\sim$ SU(2)
we stress  that invariance under  homogeneity and isotropy as well as 
gauge transformations requires that 
all components of $\phi$ are zero \cite{36}. Only for SO(N), N$>$3 
we may have  $\phi$ = (0,0,0, $\phi_1, ..., 
\phi_{N-3}$).
}}  where  $\bar \phi^{I*} = \phi^I = 0$, 
$\chi_A^I = \bar \chi_{A'}^{I*} = 0$   with a gauge group $\hat{G} = SU(2)$. 
 
\vspace{0.4cm}

An important consequence of not having scalar fields and their fermionic partners is that 
the Killing potentials $D^{(a)}$,  and related quantitites are now absent. In fact,   
if we had complex 
scalar fields, a K\"ahler manifold could be considered with metric 
$ g_{I J^*}$
on the space of $ ( \phi^I, \phi^{J^*} ) $. 
Analytic isometries that preserve the analytic structure of the 
manifold are associated with Killing vectors through \cite{34}
\begin{eqnarray}
X^{(b)} &= & X^{I (b)} ( \phi^J )~{\pt \over \pt \phi^I}~, \nonumber \\
X^{^* (b)} &= & X^{I^* (b)} ( \phi^{J^*} )~{\pt \over \pt \phi^{I^* }}~.
\label{eq:41}
\end{eqnarray}
  Killing's equation integrability 
condition is equivalent to the statement that there exist scalar functions 
 $D^{(a)}$ such that
\begin{eqnarray}
\rg_{I J^*} X^{J^* (a)} & = & i~{\pt \over \pt \phi^I}~D^{(a)}~, \nonumber \\
\rg_{I J^*} X^{I (a)} &= & - i~{\pt \over \pt \Phi^{J^*}}~D^{(a)}~. 
\label{eq:42}
\end{eqnarray}
The Killing potentials $D^{(a)}$ are defined up to constants $c^{(a)}$.
They further satisfy 
\begin{equation}
\left[ X^{I (a)} {\pt \over \pt \phi^I} + X^{I* (a)} {\pt \over \pt 
\phi^{I*}}\right]D^{(b)} = 
- f^{abc} D^{(c)}.~ 
\label{eq:43}
\end{equation}
 This fixes the constants $c^{(a)}$ for non-Abelian 
gauge groups. 
For $\hat{G}=SU(2)$ we take (see ref. \cite{34}) 
$ K = \ln(1 + \phi \bar \phi)$ and the functions 
\begin{equation}
 D^{(1)}= \half \left({{\phi + \bar \phi}\over {1 + \phi\bar\phi}}\right),
~D^{(2)}= - {i \over 2} \left({{\phi - \bar \phi}\over {1 + \phi\bar\phi}}\right),
~D^{(3)}=~- \half \left({{1 - \phi\bar\phi}\over {1 + \phi\bar\phi}}\right)
,~\label{eq:44} \end{equation}
follow. For $\phi=\ol\phi = 0$ this implies $D^{(1)}=D^{(2)}=0$ and 
$D^{(3)}=~- \half$. However, being the 
$D^{(a)}$ fixed up to  constants which are now arbitrarly, we can choose 
 $D^{(3)} = 0$ consistently.
Basically, $\phi=\ol\phi = 0$ also implies  the inexistence of Killing vectors.
Without complex scalar fields
and fermionic partners it is meaningless to discuss  K\"ahler manifolds, their 
analytical isometries and integrability conditions associated with $D^{(a)}$.

As we will explain in the following, we will  use again the 
ans\"atze (\ref{eq:1}), (\ref{eq:2}) for the tetrad and gravitinos.
Concerning the   matter fields,   the simplest  choice 
 would be to take $A^{(a)}_\mu, \lambda^{(a)}_A$ and Hermitian conjugates
as time-dependent only. However, this is not sufficient 
 in ordinary quantum cosmology with 
Yang-Mills fields.
Special ans\"atze are required\footnote{ The introduction of fermions in 
quantum cosmological models with gauge fields seemed  to require 
additional {\it non-trivial} ans\"atze for the fermionic fields \cite{38,39}.
That produced  
more restrictions  
that went beyond a simple  spatial-independency  imposed on the fermions. 
Fermions in simple  minisuperspace models have also been considered in 
ref. 
\cite{41,42,43}. 
Some questions concerning the 
(in)consistency of these models were raised in ref. \cite{41}
 and an attempt to clarify them was made in ref. \cite{42}.}
 for $A^{(a)}_\mu$ (and $\phi$ as well) 
\cite{35}-\cite{37} which depend on 
 the 
gauge group  considered. 

We will consider here  the ansatz described in \cite{35}-\cite{37} for the 
vector field $A^{(a)}_\mu$ and also employed in 
\cite{23,24}. This  ansatz is the simplest one 
that allows  vector 
fields to be present in a FRW geometry. In fact, 
only non-Abelian spin-1 fields can exist consistently within 
 a FRW background \cite{35}-\cite{37}.
  More specifically, since  the physical 
observables are to be $SO(4)$-invariant, the 
fields with gauge degrees of freedom may 
transform under $SO(4)$ if these
 transformations can be compensated by a gauge 
transformation. This is so since the physical observables 
are gauge invariant quantities. Fortunately 
there is a large class of fields satisfying the above 
conditions. These are the so-called {\it $SO(4)$--symmetric 
fields}, i.e.,  fields which are invariant up to a gauge 
transformation. Assuming a gauge group 
$\hat{G} = SO(3) \sim SU(2)$,
 the  {\it $SO(4)$--symmetric}
spin-1 field is taken to be 
\begin{equation}
{\bf A}_{\mu}(t)~\omega^{\mu}  = 
\left(
{{f(t)}\over {4}}
\varepsilon_{(a)i(b)}{\cal T}^{(a)(b)}\right)\omega^i
~.\label{eq:45}
\end{equation}
Here $\{\omega^{\mu}\}$ represents the moving coframe 
$\{\omega^{\mu}\} = \{ dt,\omega^i\}$, $\omega^i = \hat E^i_{~\hat c} dx^{\hat c}$ 
$~(i,\hat c = 1,2,3)~$ of 
one-forms, invariant under the left action of $SU(2)$ and ${\cal T}_{(a)(b)}$
are the generators of the $SU(2)$ gauge group, 
with $\tau_{(a)} = - \frac{1}{2} \eps_{(a)(b)(c)} {\cal T}_{(b)(c)}$ 
being the usual $SU(2)$  matrices.
The idea behind the ansatz (\ref{eq:45})   is to define a homorphism of the 
isotropy group $SO(3)$ to the gauge group. 
This homomorphism defines the gauge transformation
which, for the symmetric fields,
 compensates the action of a given $SO(3)$ rotation. 
Hence, the above form 
for the gauge field,  where the $A_0$ component is taken 
to be identically zero. 
None of the 
 gauge symmetries will survive: all the available
  gauge transformations are required to 
cancel out the action of a given $SO(3)$ rotation. 
Thus, we will not have in our FRW case a gauge  
constraint\footnote{{\rm 
However, in the case of larger gauge group 
some of the 
 gauge symmetries will survive. These will give rise, in the 
one-dimensional model, to local internal symmetries with 
a reduced gauge group. 
Therefore,   a gauge  
constraint can be expected to play an important role 
in such a case  and a study of such a model 
would be  interesting.}}
 $Q^{(a)}=0$.

In addition,  the ansatz (\ref{eq:45})  
 implies $A^{(a)}_\mu$ to be paramatrized by 
a single  scalar function $f(t)$. 
FRW  cosmologies with this   ansatz are totally 
equivalent to a FRW minisuperspace with an effective 
 conformally coupled 
scalar field,  but with a quartic potential instead of a 
quadratic one. 
Such minisuperspace sector {\it simplifies ~ considerably } 
any analysis of the Hamiltonian constraints \cite{35,38} 
and this constituted another  compelling argument to use ansatz (\ref{eq:45}).

We could now  take as 
fermionic partner for $A^{(a)}_\mu$ 
 a simple spin-${1 \over 2}$ field, like $\chi_A$.
This seems reasonable and in similarity  with the case   
where only scalar fields are present \cite{10,11}.
However, we will see that this choice would lead to 
some difficulties. Hence we will use the more general choice
\begin{equation}
\lambda_A^{(a)}
= 
\lambda_A^{(a)} (t)~.
\label{eq:46}
\end{equation}


Let us now address the consequences of the ans\"atze (\ref{eq:1}), 
(\ref{eq:2}), 
(\ref{eq:45}), 
(\ref{eq:46}) 
as far as the transformations 
(\ref{eq:23})-(\ref{eq:40}) are concerned. 

The tetrad (\ref{eq:1}) 
  is not affected by any gauge transformation. Moreover, 
the 
Lorentz, local coordinate and supersymmetry 
 transformations of the tetrad 
are precisely the same as in the case of 
pure N=1 supergravity. Hence, we see no reason to change the 
ansatz (\ref{eq:1})  and  the comments present in subsection 
2.1 will hold.
Namely, that the ansatz (\ref{eq:1}) is preserved under a combination of 
Lorentz, local coordinate, (obviously gauge) and \susy~ transformations. 
We should also stress that  only  the Rarita-Schwinger field is present 
in the transformations (\ref{eq:23}), (\ref{eq:28a}), (\ref{eq:32}). Thus, 
 to get a result similar to 
eq. (\ref{eq:10}) we have to employ  
  the ansatz (\ref{eq:2}) without any change.

Concerning then the  ansatz (\ref{eq:2}), we may 
neglect  the transformation (\ref{eq:37}). In fact, from $\phi = \bar\phi = 0$  
it follows that  $F^{(a)}$ can be subsequently set equal to zero. 
The  transformations (\ref{eq:29}) and (\ref{eq:33}) 
will contribute with $2D_\mu \epsilon^A$ to an expression 
equal to eq.  (\ref{eq:21}). Let us address the remaining of 
transformation  (\ref{eq:24}),  namely the term 
\[ {i\over 2} \left(g_{\mu\nu} \varepsilon^{AB} + 
C_{\mu\nu}^{~~AB}\right)\epsilon_B \lambda^{(a)C} e^\nu_{CC'} \bar\lambda^{(a)C'}
.\] 
This simplifies to 
\begin{equation}
-\frac{3i}{8} \epsilon^A \lambda^{(a)C} e_{iCC'} 
\bar\lambda^{(a)C'} - \frac{3i}{4} \epsilon_C \lambda^{(a)C} 
e^A_{~C'i} \bar\lambda^{(a)C'}~.
\label{eq:47}
\end{equation}
In order 
to achieve an expression similar to   
(\ref{eq:20}) further conditions ought to be imposed. 
Equating the first term in (\ref{eq:47}) to zero 
gives essentially the contribution of the spin-$\half$  fields to the Lorentz 
constraint. This is an interesting 
way of  analysing the Lorentz constraints when 
supermatter is present and has not been stressed previously in the current 
literature. 
We further need  to consider the term 
$\epsilon_C \lambda^{(a)C} 
e^A_{~C'i} \bar\lambda^{(a)C'}$ 
as representing a field  variable  with indices $A$ and $i$, for each 
value of $(a)$. Notice that to preserve the ansatz (\ref{eq:2}) 
  in subsection 2.1 (see also ref. \cite{11}) 
we had to require 
  $n_{AB'}\eps^B\sim \bar\psi_{B'}$ 
which is not quite $\bar\psi$. Here we have to deal with the 
$\lambda, \bar\lambda$--fields 
and a similar step is necessary. 

Moving now to the homogeneous spin-1 field, the components of 
the ansatz  (\ref{eq:45}) in the basis $\left(E^i_{\hat c} dx^{\hat c}, 
\tau_{(a)}\right)$ can be
expressed as 
\begin{equation}
A_i^{(a)} = \frac{f}{2} \delta_i^{(a)}~. 
\label{eq:48}
\end{equation}
The local coordinate and Lorentz transformations 
will correspond  to isometries and local rotations and these have been 
compensated by gauge transformations (cf. ref. \cite{35}-\cite{37} for more 
details). So,  will \susy~ transformations (\ref{eq:25}) preserve 
expressions (\ref{eq:45}) or (\ref{eq:48})? Using ans\"atze 
(\ref{eq:1}), (\ref{eq:2}), (\ref{eq:45}), (\ref{eq:46}), we get 
\begin{equation}
\delta_{(s)} A_i^{(a)} = iaE^b_i\sigma_{bAA'} \left(
\epsilon^A\bar\lambda^{A'(a)} 
-   \lambda^{A(a)}\bar\epsilon^{A'}\right)~.
\label{eq:49}
\end{equation}
Hence,  we need to impose the following 
condition\footnote{If we had chosen $\lambda^{(a)}_A = \lambda_A$ for 
any value of $(a)$ then we would not be able to obtain a 
consistent relation similar to (\ref{eq:50}). Namely, such that 
$\delta A^{(1)}_1 \sim A^{(1)}_1 $ and 
$\delta A^{(2)}_1 \sim A^{(2)}_1 = 0$.}
\begin{equation}
\left\{
\begin{array}{ccl}
\sigma_{bAA'} \left(\epsilon^A\bar\lambda^{A'(a)}
-   \lambda^{A(a)}\bar\epsilon^{A'}\right)  & = & E(t)~,~~\leftarrow
\left\{
\begin{array}{c}
(a)=b=1 \\
(a)=b=2 \\
(a)=b=3\\
\end{array}
\right.
\\
\sigma_{bAA'} \left(\epsilon^A\bar\lambda^{A'(a)}
-   \lambda^{A(a)}\bar\epsilon^{A'}\right)  & =  & 0~, ~~\leftarrow (a) \neq b 
\end{array}
\right. ~,
\label{eq:50}
\end{equation}
 where $E(t)$ is spatially independent and possibly complex, 
in order to obtain $\delta A^i_{(a)} = P_3 \left[e_{AA'\mu}, 
\psi^A_\mu,\right.$  $\left.
A^{(a)}_\mu, \lambda_A^{(a)}\right] A_{(a)}^i$ from 
(\ref{eq:45}).
From eq. (\ref{eq:50}) it follows that the preservation of the ansatz (\ref{eq:45}) will 
require $\delta A^{(a)}_i$ to include a nilpotent (``soul'') component. 
This consequence is similar to (\ref{eq:15}),  (\ref{eq:16}), (\ref{eq:21}) for the 
tetrad and gravitinos, and in accordance with  the method introduced 
in ref. \cite{11}.

As far as the $\lam-$fields are concerned, we obtain the following result 
using the ans\"atze (\ref{eq:1}), (\ref{eq:2}), (\ref{eq:45}), (\ref{eq:46}) and 
transformations (\ref{eq:26}), (\ref{eq:31}), (\ref{eq:35}), (\ref{eq:40}):
\begin{eqnarray}
 \delta \lam^{(a)}_A = 
& - & \half {\cal F}_{0i}^{(a)} e_A^{~A'i}n^B_{A'}\eps_B
 \nonumber \\
& + & \frac{i}{2} {\cal F}_{ij}^{(a)} \varepsilon_{ijk} h^{\half} n_{AA'} 
e^{BA'k} \eps_B 
\nonumber \\
 & - & \frac{i}{4} \psi_{A0}\lambda^{(a)A'} n^B_{~A'} \eps_B 
- \frac{i}{8} \psi^C_0 n_{CC'} \bar\lam^{(a)C'} \eps_A 
\nonumber \\
& -& \frac{i}{4} \bar\psi^{A'}_0 \lambda_A^{(a)} n^{B}_{~A'} \eps_B 
+ \frac{i}{8} \bar\psi_{C'0} n^{CC'} \lam^{(a)}_C \eps_A + \kappa^{abc}\zeta^b 
\lam^{(c)}_A 
\nonumber \\
& - & \frac{i}{4} \bar\psi^{A'} n_{AE'} \bar\lam^{(a)E'}n^B_{A'} \eps_B 
- \frac{i}{16}  \bar\psi_{E'}  \bar\lam^{(a)E'} \eps_A
 \nonumber \\
& -& i\bar\psi_{E'} n^{BE'} \bar\lam^{(a)A'} n_{AA'} \eps_B 
+ i\bar\psi_{E'}\bar\lam^{(a)E'}\eps_A 
\nonumber \\
& - & \frac{1}{2}\psi^F\lambda^{(a)}_A\eps_F - \frac{i}{\sqrt{2}} 
\psi^C\lam^{(a)}_C\eps_A
\nonumber \\
& + & \frac{i}{8} \psi_A  \lam_E^{(a)}  \eps^A 
- \frac{i}{16} \psi_C  \lam^{(a)C} \eps_A 
\nonumber \\
& + & 2\eps_A \lam^{(a)}_B \psi^B + \eps_C\psi^C\lambda^{(a)}_A
~,
\label{eq:51}
\end{eqnarray}
where
\begin{eqnarray}
{\cal F}_{0i}^{(a)} & = & \dot{f}\delta^{(a)}_i ~,\nonumber \\
{\cal F}_{ij}^{(a)} & = & \frac{1}{4}(2f - f^2) \varepsilon_{ij(a)}~, 
\label{eq:52}
\end{eqnarray}
and the following relation were employed
\begin{eqnarray}
\eps_{ijk} e^i_{AA'} e^j_{BB'} e^k_{CC'} =& 
-& ih^\half \eps_{AB} n_{EA'} \eps_{D'B'} \left(
-\eps_C^{~E} \eps_{A'}^{~D'} + n_{CC'} n^{ED'}
\right)
\nonumber \\
& + & ih^\half \eps_{A'B'} n_{BE'} \eps_{DA} 
\left(
\eps_C^{~D} \eps_{C'}^{~E'} 
+ n_{CC'} n^{DE'} \right) ~.
\label{eq:53}
\end{eqnarray}
Using the bi-spinorial relation (\ref{eq:13}) and eq.
(\ref{eq:50}), 
the last sixth  terms in eq. 
(\ref{eq:51}) may be put in a more suitable form in order to 
get 
$\delta\lam^A_{(a)} = P_4 \left[e_{AA'\mu}, \psi^A_\mu, 
A^{(a)}_\mu, \lambda_A^{(a)}\right] \lam^A_{(a)}$. 
This would require that the remaining terms 
(including the first to the 12th term in (\ref{eq:51})) 
 to satisfy a further 
condition equated to zero.

The above results concerning the transformations
(\ref{eq:23})-(\ref{eq:40}) imply that the ans\"atze for the physical 
variables are   consistent 
with a 
   FRW geometry and Lorentz, gauge and supersymmetry 
transformations. However, some restrictions (see eq. 
(\ref{eq:16}), (\ref{eq:20})--\ref{eq:22}), (\ref{eq:47}), 
(\ref{eq:50}), (\ref{eq:51})) had to be imposed. It is reasonable 
to consider that such limitations would 
affect the possible spectrum of quantum solutions. In fact, 
  it does and the solutions present in section 3 and ref. \cite{23,24} 
constitute just the expected consequence of these restrictions.

The purpose of section 3 is precisely  to show 
both 
the usefulness and limitations of 
 the ans\"atze for $e^a_\mu, \psi^A_\mu$, 
$ A^{(a)}_\mu, \lambda_A^{(a)}$ 
 in deriving interesting results.   We will reproduce previous results 
present in the literature regarding  {\it non-}supersymmetric quantum 
cosmologies with Yang-Mills fields \cite{35}. Being 
N=1 supergravity a square-root of gravity, our results are thus 
particularly supportive. Hence, our model 
could  be regarded as valued  approach   in 
persuing a locally \susy~FRW with the {\em most} 
 general gauge supermatter \cite{23,24}.
Nevertheless, the solutions obtained here correspond to 
just a subset of the ones in ref. \cite{35}. It seems then that
our ans\"atze are {\em not} general enough 
 in spite 
of their simplicity and adequacy. Hence, the spectrum of solutions 
is {\it incomplete} or drastically limited ($\Psi = 0 $ in ref. 
\cite{23,24}).

\section{ Quantum constraints and solutions}

\indent

Let us then  solve explicitly the corresponding quantum Lorentz and 
supersymmetry constraints\footnote{The use of 
these constraints signifies  that supersymmetry and Lorentz invariance 
is a feature of the reduced FRW model but subject to the 
(restrictive) conditions(\ref{eq:16}), (\ref{eq:20})--\ref{eq:22}), (\ref{eq:47}), 
(\ref{eq:50}), (\ref{eq:51}).}.
 First we need to redefine the  fermionic fields,  $ \psi_{A} $ and 
$\lambda^{(a)}_{A}$ 
in order to simplify the Dirac 
brackets \cite{2,11} following  the steps described in \cite{24}.


 For the   $ \psi_{A}$-field we introduce, 

\begin{equation} \hat \psi_{A} = 
{\sqrt{3} \over 2^{1 \over 4}} \sigma a^{3 \over 2} \psi_{A}~, ~
\hat{\bar \psi}_{A'} = {\sqrt{3} \over 2^{1 \over 4}} \sigma a^{3 \over 2} \bar \psi_{A'} ~, 
\label{eq:54}
\end{equation}
where the conjugate momenta are 

\begin{equation} \pi_{\hat{ \psi}_{A}} = in_{AA'} \hat{\bar \psi}^{A'} ~,~
\pi_{\hat{\bar  \psi}_{A'}} = in_{AA'} \hat \psi^{A}~.  
\label{eq:58}
\end{equation}
The  Dirac brackets then become

\begin{equation} [\hat \psi_{A} , \hat{\bar \psi}_{A'}]_{D} = in_{AA'}~.  
\label{eq:59}
\end{equation}
Similarly for the $\lambda_A^{(a)}
$ field

\begin{equation} \hat\lambda^{(a)}_{~A} = 
{\sigma a^{ 3 \over 2} \over 2^{ 1 \over 4} }\lambda^{(a)}_{~A}, ~~
\hat{\bar \lambda}^{(a)}_{~A'} = 
{\sigma a^{ 3 \over 2} \over 2^{ 1 \over 4} }\bar \lambda^{(a)}_{~A'}~, 
\label{eq:60} 
\end{equation}
giving
\begin{equation}  \pi_{\hat{ \lambda}^{(a)}_{~A}} = -in_{AA'} \hat{\bar \lambda }^{(a)A'} ~,~
\pi_{\hat{\bar \lambda}^{(a)}_{~A'}} = -in_{AA'} \hat \lambda^{(a)A}~ , \label{eq:61}
\end{equation}
with
\begin{equation}  [\hat \lambda^{(a)}_{~A}, \hat{\bar \lambda}^{(a)}_{~A'}]_{D} = - i \delta^{ab} n_{AA'} ~.
\label{eq:62}
\end{equation}
Furthermore,
\begin{equation} [a , \pi_{a}]_{D} = 1~,  ~[f, \pi_{f}]_{D} = 1,  
\label{eq:63}\end{equation}
and the rest of the brackets are zero.

It is simpler to describe the theory using only (say) unprimed spinors, and, to this end, we define
\begin{equation} \bar \psi_{A} = 2 n_{A}^{~B'} \bar \psi_{B'}~, ~
 \bar \lambda^{(a)}_{~A} = 2 n_{A}^{~B'}
 \bar \lambda^{(a)}_{~B'} ~,
\label{eq:64}
\end{equation}
with which the new Dirac brackets are 
\begin{equation}  
 [\psi_{A}, \bar \psi_{B}]_{D} = i \epsilon_{AB} ~, [\lambda^{(a)}_{~A}, \bar \lambda^{(a)}_{~A'}]_{D} =
 - i \delta^{ab} \epsilon _{AB}~. \label{eq:65} 
 \end{equation}
The rest of the brackets remain unchanged. 
Quantum mechanically, one replaces the Dirac brackets by  anti-commutators if both arguments are odd 
(O) or commutators if 
otherwise (E): 

\begin{equation} [E_{1} , E_{2}] = i [E_{1} , E_{2}]_{D} ~,~ [O , E] = i [O , E]_{D} ~,~ \{O_{1} , O_{2}\} = i [O_{1} , O_{2}]_{D} ~.
\label{eq:66}
\end{equation}
Here, we take units with $\hbar = 1 $. The only non-zero (anti-)commutator relations are:
\begin{equation} \{\lambda^{(a)}_{~A}, \lambda^{(b)}_{~B} \} = \delta^{ab} \epsilon_{AB} ~, ~
\{\psi_{A} , \bar \psi_{B}\}  = - \epsilon_{AB}~,~
 [a , \pi_{a}]   = [f, \pi_{f}] = i  ~.
\label{eq:67} \end{equation}
We chose $ \left
(\bar\lam_{A'}^{(a)},\psi_{A} , a, f\right) $ to be the coordinates of the configuration 
space, and 
$ \left(\lam_{A}^{(a)},\bar \psi_{A}, \pi_a, \pi_f\right)$ to be the momentum operators in this representation.
Hence

\begin{equation}  \lambda^{a}_{~A} \rightarrow   -{  \partial   \over   \partial   \bar \lambda^{(a)A}} ~,~
  \bar \psi_{A} \rightarrow {  \partial   \over   \partial   \psi^{A}}~,~
  \pi_{a}  \rightarrow  {  \partial   \over   \partial   a} , ~  \pi_{f} \rightarrow -i {  \partial   \over   \partial   f}~.  
\label{eq:68}
\end{equation}

Following the ordering used in ref. \cite{5}, 
 we put all the fermionic derivatives in  $S_{A}$ on the right.
 In $ \bar S_{A} $, all the fermonic
derivatives are on the left. Implementing all these redefinitions, the 
 supersymmetry constraints  have the differential operator form

\begin{eqnarray} 
S_{A} & = &  
 - {1 \over 2 \sqrt{6}} a \psi_{A} {  \partial   
\over   \partial   a}   
 -   \sqrt{3 \over 2} \sigma^{2}a^{2} \psi_{A} \nonumber \\
& - & {1 \over 8 \sqrt{6}} \psi_{B} \psi^{B} {  \partial   \over   \partial   \psi^{A}} 
 - {1 \over 4 \sqrt{6}} \psi^{C} 
\bar \lambda^{(a)}_{~C} {  \partial   \over   \partial   \bar \lambda^{(a)A}}
 \nonumber \\
  & + & {1 \over 3 \sqrt{6}} \sigma^{a}_{~AB'} \sigma^{bCC'} n_{D}^{~B'} n^{B}_{~C'} \bar \lambda^{(a)D} \psi_{C} {  \partial   \over   \partial   
 \bar \lambda^{(b)B}}  \nonumber \\
&+ & {1 \over 6 \sqrt{6}} \sigma^{a}_{~AB'} \sigma^{bBA'}
 n_{D}^{~B'} n^{E}_{~A'} \bar \lambda^{(a)D} \bar \lambda^{(b)}_{~B}
{  \partial   \over   \partial   \psi^{E}}  \nonumber \\
& -& { 1 \over 2 \sqrt{6}} \psi_{A} \bar \lambda^{(a)C} {  \partial   \over 
\partial 
\bar \lambda^{(a)C}} + {3 \over 8 \sqrt{6}} \bar 
\lambda^{a}_{~A} \lambda^{(a)C} {  \partial  
 \over   \partial   \psi^{C}}  
\nonumber \\   
  & + & \sigma^{a}_{~AA'} n^{BA'}  \bar \lambda^{(a)}_{~B} \left( -{ \sqrt{2} \over 3}  {   \partial   \over   \partial   f} + 
{1 \over 8 \sqrt{2}} (1 -(f-1)^{2}) \sigma^{2}  \right)  
\label{eq:69}
\end{eqnarray} 
and 
\begin{eqnarray}
 \bar S_{A} & = &  {1 \over 2 \sqrt{6}} a {\pt \over \pt a} {\pt \over \pt \psi^{A}} 
 - \sqrt{3 \over 2} \sigma^{2}a^{2} {\pt \over  \pt \psi^{A}} 
 \nonumber \\
&- & {1 \over 8 \sqrt{6}} \varepsilon^{BC} { \pt \over \pt \psi^{B}} {\pt \over \pt \psi^{C}} \psi_{A} + {1 \over 4 \sqrt{6}} \varepsilon^{BC} {\pt \over \pt \psi^{B}} 
{\pt \over \pt \bar \lambda^{(a)C}} \bar \lambda^{(a)}_{~A}
 \nonumber \\
&+ & {1 \over 3 \sqrt{6}}  \sigma^{aB}_{~~A'} 
\sigma^{bCC'} n_{A}^{~A'} n^{D}_{~C'} {\pt \over \pt \psi^{D}}
{\pt \over \pt \bar \lambda^{(a)B}} \bar \lambda^{(b)}_{~C} \nonumber \\
&+ &  {1\over 6 \sqrt{6}}  \sigma^{aB}_{~~A'} 
\sigma_{D}^{~bB'} n_{A}^{~A'} n^{C}_{~B'} {\pt \over \pt \bar \lambda^{(a)B}}
{\pt \over \pt \bar \lambda^{(b)C}} \psi^{D} \nonumber \\
&+ &  { 1 \over 2 \sqrt{6}} {\pt \over \pt\psi^{A}} {\pt 
\over \pt \bar \lambda^{(a)B}} \bar \lambda^{(a)B} 
+{ 3 \over 8\sqrt{6}} {\pt \over \pt \bar \lambda^{(a)B}} 
{\pt \over \pt \bar \lambda^{(a)A}} \psi^{B} \nonumber \\
&+&   n_{A}^{~A'} \sigma^{aB}_{~~A'} \left( { 2 \sqrt{2} \over 3} 
{\pt \over \pt f} + { 1 \over 4 \sqrt{2}} (1 - (f-1)^{2})
\sigma^{2} \right) {\pt \over \pt \bar \lambda^{(a)B} }. 
\label{eq:70}
\end{eqnarray}

When matter fields are taken into account the
generalisation of the $ J_{AB} $
constraint  is :
\begin{equation} J_{AB} = \psi_{(A} \bar\psi^{B'}n_{B)B'} - 
  \lambda^{(a)}_{(A} \bar\lambda^{(a)B'}n_{B)B'} =  0~. 
 \label{eq:76a}
 \end{equation}
 The Lorentz constraint $ J_{AB} $ implies that a physical 
wave function should be a  Lorentz scalar. 
 We can easily see that the most general form of the wave function 

\begin{eqnarray} 
  \Psi = A & + & B \psi^{C} \psi_{C}  
 + d_{a}  \lambda^{(a)C} \psi_{C} + c_{ab} \bar \lambda^{(a)C}  \bar \lambda^{(b)}_{~C}
+ e_{ab}  \bar \lambda^{(a)C}  \bar \lambda^{(b)}_{~C} \psi^{D} \psi_{D}  
\nonumber \\ 
 & + & 
 c_{abc}  \bar \lambda^{(a)C}  \bar \lambda^{(b)}_{~C} \bar \lambda^{(c)D} \psi_{D}   
  + c_{abcd} \bar \lambda^{(a)C}  \bar \lambda^{(b)}_{~C}
\bar \lambda^{(c)D}  \bar \lambda^{(d)}_{~D} + d_{abcd}
\bar \lambda^{(a)C}  \bar \lambda^{(b)}_{~C} \bar \lambda^{(c)D}  \bar \lambda^{(d)}_{~D}  \psi^{E} \psi_{E}   \nonumber \\ 
& + &
  \mu_{1}  \bar \lambda^{(2)C}  \bar \lambda^{(2)}_{~C} \bar \lambda^{(3)D} \bar \lambda^{(3)}_{~D} \bar \lambda^{(1)E} \psi_{E} 
\nonumber \\
&+ & \mu_{2}  \bar \lambda^{(1)C}  \bar \lambda^{(1)}_{~C} \bar \lambda^{(3)D} \bar \lambda^{(3)}_{~D} \bar \lambda^{(2)E} \psi_{E}
+ \mu_{3}  \bar \lambda^{(1)C}  \bar \lambda^{(1)}_{~C} \bar \lambda^{(2)D} \bar \lambda^{(2)}_{~D} \bar \lambda^{(3)E} \psi_{E}   \nonumber \\
& + & F 
\bar \lambda^{(1)C}  \bar \lambda^{(1)}_{~C} \bar \lambda^{(2)D} \bar \lambda^{(2)}_{~D} \bar \lambda^{(3)E} \bar \lambda^{(3)}_{~E} 
  +   G  \bar \lambda^{(1)C}  \bar \lambda^{(1)}_{~C} \bar \lambda^{(2)D} \bar \lambda^{(2)}_{~D} \bar \lambda^{(3)E} \bar \lambda^{(3)}_{~E}
\psi^{F} \psi_{F}~.
  \label{eq:71}
 \end{eqnarray}
where $A$, $B$,...,$G$  are functions of $a$,  $f$  only.  
This Ansatz contains all allowed combinations of the fermionic fields and 
 is the most general Lorentz invariant function we can write down.

The next step is to solve the supersymmetry constraints $ S_{A} \Psi = 0 $ and $ \bar S_{A'} \Psi = 0 $. 
Since the wave function (\ref{eq:71}) is of  even order in fermionic variables 
and 
stops at order $8$, the 
 equations $S_{A} \Psi = 0$ and $\bar S_{A} \Psi = 0$ will be of 
   odd order in fermionic variables and stop at order $7$.
Hence we will get ten  equations from $S_A \Psi = 0$ and another 
ten equations from $\ol S_A \Psi = 0$. From $S_A \Psi = 0$  we obtain

\begin{equation}      
-{a \over {2\sqrt{6}}} {\partial A \over \partial a} - \sqrt{{3\over 2}} \sigma^2 a^2 A = 0, \label{eq:72}     
\end{equation}          
\begin{equation}      
-{\sqrt{2} \over 3} {\partial A \over \partial f} + {1 \over {8\sqrt{2}}} \left[1 - (f - 1)^2\right] \sigma^2~A =0~.  \label{eq:73}   
\end{equation}          
These equations  correspond, respectively, to terms linear in $\psi_A, 
\bar\lambda^{(a)}_A$. Eq. (\ref{eq:72}) and (\ref{eq:73}) give the       
dependence of $A$ on $a$ and $f$, respectively. 
Solving these equations leads to       
$A = \hat A (a) \tilde A (a)$ as         
\begin{equation}      
A = \tilde A (f) e^{-3 \sigma^2 a^2},      
\label{eq:74}\end{equation}          
\begin{equation}      
A = \hat{A}(a)  e^{{3\over 16}\sigma^2 \left(-{f^3 \over 3} + f^2\right)},      
\label{eq:75} 
\end{equation}      
A similar relation exists for the
 $\bar S_A \Psi = 0$ equations, which from the       
$\psi^A\lambda^{(1)}_E\lambda^{(1)E} \lambda^{(2)}_E\lambda^{(2)E} 
\lambda^{(3)}_E\lambda^{(3)E} $ term in $\Psi$ give for $G = \hat G (a) \tilde G (f)$        
\begin{equation}      
G = \tilde G (f) e^{3 \sigma^2 a^2},     
\label{eq:76} 
\end{equation}          
\begin{equation}      
G = \hat{G}(a)  e^{{3\over 16}\sigma^2 \left({f^3 \over 3} - f^2\right)}.      
\label{eq:77}
\end{equation}        
We notice that 
in our case study, differently to the case of ref. \cite{5,11,22}-\cite{28}, we are 
indeed 
allowed to completely determine the dependence of $A$ and $G$ with respect to 
$a$ and $f$.

The solution (\ref{eq:76}), (\ref{eq:77}) 
is included   in the 
  Hartle-Hawking 
(no-boundary) 
solutions of ref. \cite{35}. In fact, we basically recover solution (3.8a)   in ref. 
\cite{35} if we replace 
$f \rightarrow f + 1$. 
As it can be checked, this procedure 
constitutes the rightful choice  
according to the definitions employed in \cite{36} for $A^{(a)}_\mu$. 
Solution  (\ref{eq:76}), (\ref{eq:77})
is also associated with an anti-self-dual solution of the Euclidianized equations 
of motion (cf. ref. \cite{35,35a}). 
However, it is relevant to emphasize that   {\em not all} 
 the solutions present in 
\cite{35} can be recovered here. In particular, the Gaussian wave function 
(\ref{eq:76}), (\ref{eq:77}), peaked around $f=1$ 
(after implementing the above transformation), 
represents only one of the components of the 
wave function in ref. \cite{35}. The   wave 
function in ref. 
\cite{35}
 is peaked around the {\it two} 
 minima of the corresponding quartic potential. 
In our model, 
the  potential terms 
 correspond  to a ``square-root'' of the potential present in \cite{35}.

Solution (\ref{eq:74}), (\ref{eq:75})  has the features of a  
(Hawking-Page)
wormhole solution for Yang-Mills fields 
\cite{33,35a}, 
which nevertheless has not yet  been found in  ordinary quantum cosmology.
However, in spite of  
(\ref{eq:74}), (\ref{eq:75})
being regular for $a \rightarrow 0$ and damped for 
$a \rightarrow \infty$, it may not be well behaved when $f \rightarrow -\infty$.

The equations obtained from the cubic and 5-order fermionic terms in 
$S_A\Psi = 0$ and $\bar S_A \Psi = 0$
 can be dealt with by multiplying them  by 
$n_{EE'}$ and   using the relation $n_{EE'} n^{EA'} = {1 \over 2} \epsilon_{E'}^{~A'}$ 
Notice  that  the $\sigma_a$ 
matrices are linear independent  and are orthogonal to the $n$ matrix.
We would see that such equations  provide the 
  $a, f$-dependence of the remaining terms in $\Psi$.
It is important to point out 
that the dependence 
of the coefficients in $\Psi$ corresponding to cubic fermionic 
terms 
on $a$ and $\phi,\ol\phi$ is 
{\it mixed }
 throughout several equations \cite{5,11}.  However, 
in the present  FRW minisuperspace with 
vector fields, the analogous dependence in $a, f$ occurs in 
{\it separate} equations. 
The  equations for cubic and 5-order fermionic terms further 
  imply that  any possible solutions
are neither the Hartle-Hawking or a wormhole state. In fact,  
we would get $d_{(a)} \sim a^5 \hat d_{(a)}(a) \ti d_{(a)} (f)$ and similar 
expressions for the other coefficients in $\Psi$,
 with a prefactor $a^n$, $n \neq 0$. This 
behaviour has also been found in \cite{5}.  Hence, from their $a$-dependence equations
these cannot be 
either a Hartle-Hawking or wormhole state. They correspond to 
other type of solutions which could be obtained from the corresponding Wheeler-DeWitt 
equation but with completely 
different boundary conditions.  

Finally, it is worthy to notice that 
the    Dirac bracket of the  supersymmetry constraints (\ref{eq:69}), (\ref{eq:70})  
induces an expression   whose bosonic sector corresponds 
to  the ({\it decoupled}) 
gravitational and vector field components  of the Hamiltonian constraint  in 
 ref. \cite{35}. 
This fact supports a relation between the   
ans\"atze (\ref{eq:45}), (\ref{eq:46}) and solutions 
(\ref{eq:76}), (\ref{eq:77}), within the context of 
N=1 supergravity being a square-root of gravity \cite{1}.

\section{ Discussions and Conclusions}

\indent

Summarizing our work, 
 we  considered the canonical formulation 
of  the more general theory of $ N = 1 $
 supergravity with supermatter \cite{28,34}
subject  to a $k=+1$  FRW geometry. 
Ans\"atze for  the gravitational and 
 gravitino fields, the gauge vector field $A^{a}_{\mu}$ and corresponding fermionic partners were then introduced. 
We  set the scalar fields and their 
supersymmetric partners equal to zero. Our purpose was to initiate a discussion 
on the  main result of ref. \cite{23,24}: 
the only allowed solution was $\Psi =0$.

Concerning  the ans\"atze employed here ({and also in ref. \cite{23,24}) 
for the field 
variables, the following points are  in order.
It was clear  from transformations (22)-(40) that the ansatz 
for the $k=+1$ FRW tetrad would have to be identical, either in 
subsection 2.1 (pure N=1 supergravity)  or subsection 2.2 
(N=1 supergravity with gauged supermatter). 
A  consistent ansatz was 
also required for the gravitino fields
in order for the tetrad ansatz (1) to be preserved under \susy~, 
Lorentz and local coordinate transformations. From the expression of 
$\delta e^{AA'}_{~~i}$ 
in the presence of  supermatter  
it was straightforward to conclude 
that 
the ansatz for $\psi^A_i, \bar\psi^{A'}_i$ ought to be 
the same as in the pure N=1 case (see ref. \cite{11}). 
In addition, both $\delta e^{AA'}_{~~i}, \delta \psi^A_i, \delta\bar\psi^{A'}_i$ 
(either in the pure or supermatter case) will 
include nilpotent (``soul'') components \cite{nilp} (see subsections 2.1, 2.2 and 
ref. \cite{11}).

In  addressing the  variation of 
$\delta\psi^A_i$ using  eq. (2) we
 had to combine \susy~, Lorentz and local coordinate transformations. 
Spin-$\frac{1}{2}$ fields $\left(
\lambda^{(a)}_A, \bar\lambda^{(a)}_{A'}
\right)$
were now present in $\delta_{(s)}\psi^A_i$. For 
$\delta\psi^A_i \sim \psi^A_i$ to hold, further conditions were 
imposed 
on the gravitino and $\left(
\lambda^{(a)}_A, \bar\lambda^{(a)}_{A'}
\right)$
fields. One consequence was to clarify the contribution 
of the $\left(
\lambda^{(a)}_A, \bar\lambda^{(a)}_{A'}
\right)$
fields to the Lorentz constraints.

With respect to the vector field $
A^{(a)}_\mu$, we   chosed a specific  ansatz 
to   simplify our calculations and 
 already used in ordinary quantum cosmology with 
some sucess \cite{35}-\cite{37}.  
Ansatz (45) fulfilled 
such conditions. In fact,   
it  induces FRW minisuperspaces with an 
{\it effective}  conformally coupled scalar field with a 
quartic potential. Eq. (45) corresponds to a {\it symmetric} 
vector field, i.e., such that the action of isometries and 
local rotations are compensated by  a gauge transformation.
Furthermore, it has the simplest possible form for non-trivial   non-Abelian 
vector fields 
and thus be accomodated in a FRW 
geometry.
Our results showed  that 
\susy~invariance 
could be achieved if   condition 
(50) was imposed. 
This 
further implied that $\delta A^{(a)}_\mu$ involved a ``soul'' \cite{nilp} component, 
similarly to what occurred
for the tetrad and gravitinos (see ref. \cite{11}). 
 
When addressing $\delta 
\lambda^{(a)}_A$, we were   able to obtain  
$\delta 
\lambda^{(a)}_A \sim 
\lambda^{(a)}_A$
if further (restrictive) conditions were imposed. 
Overall, accommodating  supersymmetry with homogeneity {\em and} 
isotropy requires specific conditions to be imposed on the 
physical fields. Expressions (\ref{eq:16})--(\ref{eq:22}) are just some of them. 
When a vector supermultiplet is included, further and more {\em severe} 
restrictions are required. The inclusion of 
more (and different) fields and  further restricitions will 
affect drastically the possible solutions. In particular, 
few or no non-trivial solutions would be derived in spite of  the 
ans\"atze being  physically adequate but {\em not} general enough.

  The purpose of section 3   was precisely  
to show that interesting but limited 
 physical features could   be derived
from our ans\"atze. A wave function constructed in the way mentioned 
in the previous section accomodates naturally the expectation that the early 
universe --- earlier than an inflationary stage --- might be dominated 
by radiation and associated fermionic fields. Our results constitute an 
approach towards such a supersymmetric scenario.

After a dimensional reduction, we derived the   constraints 
for our one-dimensional model and  solved the Lorentz and supersymmetry constraints.
We then obtained non-trivial solutions. We found expressions 
that  
 can be interpreted as corresponding  to a wormhole (Hawking-Page) 
 and 
(Hartle-Hawking) no-boundary solutions, respectively.

These results   were quite supportive. Namely,  the 
Hartle-Hawking solution found here corresponded to a component 
of the set of  solutions obtained  from 
a Wheeler-DeWitt equation in 
non-supersymmetric 
 quantum cosmology \cite{35} . 
That is consistent 
with our expectations, since N=1 supergravity is a   square root of 
 gravity. 
Moreover, the 
   Dirac bracket of the  supersymmetry constraints (\ref{eq:69}), (\ref{eq:70})  
induces an expression  whose bosonic sector is the ({\it decoupled}) 
gravitational and vector field components  of the Hamiltonian constraint  in 
 ref. \cite{35}.

As far as the problem of the null result in ref. \cite{23,24} is 
concerned, we hope our results may provide a new perspective on this 
issue. In the least, we know from the present paper that physical states 
in FRW models with gauged fields obtained from N=1 supergravity with 
supermatter indeed exist. Physical states also exit when solely scalar 
multiplets are concerned \cite{31}. Thus, we could expect to merge both 
situations   and hopefully obtain non-trivial 
states. It should be noticed however that so far {\it no} analytical solution 
has been found in non-supersymmetric FRW quantum cosmologies with 
vector {\it and } scalar fields. 

We could speculate that $\Psi = 0$ in ref. \cite{23,24} would possibly be a 
consequence of the ans\"atze for the physical variables. More precisely, that in spite
of their simplicity the employed ans\"atze were not the {\it more general} ones. 
Hence, only a very particular solution ($\Psi=0$) was possible. Do notice as well 
that the solutions (\ref{eq:76}), (\ref{eq:77}) in section 3 
constitute only {\it part} of the set present in \cite{35}. Hence, it would be tempting 
 to relate the incompleteness of the solutions 
(\ref{eq:76}), (\ref{eq:77})  (with respect to ref. \cite{35})  
to the ans\"atze (\ref{eq:45}), (\ref{eq:46}) being also {\it incomplete}. 
A more general  ans\"atze  (where supersymmetry, gauge, Lorentz and local coordinate
 transformations  are accomodatted {\it less} restrictively --- 
see eq. (\ref{eq:16}), (\ref{eq:20})--(\ref{eq:22}), 
(\ref{eq:47}), (\ref{eq:50}), (\ref{eq:51})) could  allow for a larger 
spectrum of solutions.

 However, 
we must  stress  that the absence of $\phi,\ol\phi$ and their 
fermionic 
partners played an important role as enhancing  $D^{(a)}$ to be set to zero.
The presence of the Killing potentials in the supersymmetry constraints 
of \cite{23,24} constitute a relevant  element in deriving $\Psi=0$. 
We hope to address all these issues in a future investigation. 

\vspace{0.4cm}

Supersymmetric quantum cosmology certainly constitutes an active 
and challenging subject for further research (see ref. \cite{31} for a 
review). Some problems that 
remain to be confronted 
 have been raised 
throughout this paper. 
Other interesting issues   which remain 
open and 
we are also aiming to address are the following: 
\begin{description}
\item [a)] Obtain conserved currents from $\Psi$, 
as consequence of the Dirac-like structure of the supersymmetry 
constraints \cite{newnew};
\item[b)] Test the validity of minisuperspace 
approximation in supersymmetric quantum 
cosmology;
\item [c)] Perform the  canonical quantization of minisuperspaces and 
black-holes in N=2,3 supergravities.
\end{description}

\vspace{1cm}
 
{\large\bf ACKNOWLEDGEMENTS}

\vspace{0.3cm}

The author is  grateful to 
A.D.Y. Cheng and S.W. Hawking  for pleasant conversations 
and for sharing their points of view and to O. Bertolami for useful 
comments and suggestions.
Early motivation from discussions with  O. Obr\'egon 
and 
 R. Graham   
are also acknowledged.  
 This work was supported by  
JNICT/PRAXIS XXI Fellowship BPD/6095/95.


\begin{thebibliography}{99}

 
 

\bibitem{1}  C. Teitelboim, Phys.~Rev.~Lett.~{\bf 38}, 1106 (1977).

\bibitem{2}  M. Pilati, Nuc. Phys. B {\bf 132}, 138 (1978).


\bibitem{3}   P.D. D'Eath, Phys.~Rev.~D {\bf 29}, 2199 (1984).



\bibitem{5} L.J. Alty, P.D. D'Eath and H.F. Dowker, Phys.~Rev.~D {\bf 46}, 4402
(1992).


\bibitem{7}  P.D. D'Eath, S.W. Hawking and O. Obreg\'on, Phys.~Lett.~{\bf 300}B, 44
(1993).

\bibitem{8}   P.D. D'Eath, Phys.~Rev.~D {\bf 48}, 713 (1993).

\bibitem{9}   R. Graham and H. Luckock, Phys. Rev. D  
{\bf 49}, R4981 (1994).




\bibitem{10}   P.D. D'Eath and D.I. Hughes, Phys.~Lett.~{\bf 214}B, 498 (1988).

\bibitem{11}   P.D. D'Eath and D.I. Hughes, Nucl.~Phys.~B {\bf 378}, 381 (1992).


\bibitem{12}   M. Asano, M. Tanimoto and N. Yoshino, Phys.~Lett.~{\bf 314}B, 303 (1993).

\bibitem{13} H. Luckock and C. Oliwa, 
Phys. Rev. {\bf D51} (1995) 5883.
 
\bibitem{14}  R. Graham and A. Csord\'as, {\it 
Nontrivial fermion states in supersymmetric minisuperspace},
in: 
Proceedings of the First Mexican School in Gravitation 
and Mathematical Physics, 
Guanajuato, Mexico, December 12-16, 1994 (gr-qc/9503054 );

 


\bibitem {15} R. Graham and A. Csord\'as, Phys. Rev. Lett. {\bf 74} (1995) 4926.





\bibitem{16} P.D. D'Eath, Phys. Lett. B{\bf 320}, 20 (1994).

\bibitem{17} A.D.Y. Cheng, P.D. D'Eath and 
P.R.L.V. Moniz, Phys. Rev. D{\bf 49} (1994) 5246.


\bibitem{18} A.D.Y. Cheng, P.D. D'Eath and 
P.R.L.V. Moniz,
{\rm Gravitation and Cosmology} 
{\bf 1} (1995) 12


\bibitem{19} R. Graham and A. Csord\'as, Phys. Rev. {\bf D52} (1995) 5653


\bibitem{20}  A.D.Y. Cheng, P.D. D'Eath and 
P.R.L.V. Moniz, 
~DAMTP-Report February R94/13,  

\bibitem{21}  A.D.Y. Cheng, P.D. D'Eath and 
P.R.L.V. Moniz,
{\rm Gravitation and Cosmology} 
{\bf 1} (1995) 1

 \bibitem{22} A.D.Y. Cheng and P.R.L.V. Moniz, 
Int. J. Mod. Phys. {\bf D4} (1995) 189
 




\bibitem{23}  A.D.Y. Cheng, P.D. D'Eath and P.R.L.V. Moniz, 
{\it Quantization of a Friedmann-Robertson-Walker model} {\it in N=1 Supergravity   with 
Gauged} {\it  Supermatter},  
in: Proceedings of the 1st Mexican School in Gravitation, Guanajuato, Mexico
December 12-16 1994,  gr-qc/9503009.

  
\bibitem{24}  A.D.Y. Cheng, P.D. D'Eath and 
P.R.L.V. Moniz, 
Class. Quantum Grav. {\bf 12} (1995) 1343-1353
 


\bibitem{25} P. Moniz, {\it The Case of the Missing Wormhole State}, in:   
Proceedings of the VI Moskow International Quantum Gravity Seminar, Moskow, Russia, 
12-19 June 1995,  to be published 
by World Scientific, DAMTP report R95/19, gr-qc/9506042, 

\bibitem{26} P. Moniz,  
Gen. Rel. Grav.  {\bf  28}  (1996) 97 

\bibitem{27} P. Moniz, {\it Back to Basics? or How can supersymmetry be used 
in a simple quantum cosmological model}, 
in: Proocedings of the 1st Mexican School in Gravitation, Guanajuato, Mexico
December 12-16 1994,  DAMTP report R95/20, gr-qc/9505002

\bibitem{28} P. Moniz,  {\it Quantization of the Bianchi type-IX model 
in N=1 Supergravity in the presence of 
 Supermatter}, DAMTP Report R95/21 , gr-qc/9505048,   to be published in  
 International Journal of Modern Physics {\bf A} Vol. 11 No.6 (1996) 

 
\bibitem{29}A.D.Y. Cheng and P. Moniz, {\it Quantum Bianchi Models in N=2 Supergravity with Global O(2) Internal Symmetry}
in: Proceedings of the VI Moskow International Quantum Gravity Seminar, Moskow, Russia, 
12-19 June 1995,  to be published 
by World Scientific, DAMTP report; 

\bibitem{30}  A.D.Y. Cheng and P. Moniz, {\it Canonical Quantization of 
  Bianchi Class A Models in N=2 Supergravity}, 
-- accepted for publication in Modern Phys. Lett. {\bf A} (1996).

\bibitem{31} P.V. Moniz, {\it  Supersymmetric Quantum Cosmology --- 
Shaken not Stirred}, 
  Int. J. Mod. Physics {\bf A} (invited review), 
DAMTP report  R95/53, gr-qc/9604025


\bibitem{4} G. Esposito, {\it Quantum Gravity, Quantum Cosmology and Lorentzian Geometries}, 
Sprin\-ger Verlag (Berlin, 1993) and references therein.


\bibitem{6} S.W. Hawking, Phys. Rev. D{\bf 37} 904 (1988).

\bibitem{32} J.B. Hartle and S.W. Hawking, Phys.~Rev.~D {\bf 28}, 2960 (1983).


\bibitem{33}   S.W. Hawking and D.N. Page, Phys.~Rev.~D {\bf 42}, 2655 (1990).

 

\bibitem{34}    J. Wess and J. Bagger, {\it Supersymmetry and Supergravity},
2nd.~ed. (Princeton University Press, 1992).


\bibitem{35}    O. Bertolami and J.M. Mour\~ao, Class. Quantum Grav. {\bf 8} (1991) 1271; 

\bibitem{35b} O. Bertolami and P.V. Moniz, Nuc. Phys. {\bf B439} (1995) 259  

\bibitem{35a} O. Bertolami, J. Mour\~ao, R. Picken and I. Volobujev, 
Int. J. Mod. Phys. {\bf A6} (1991) 4149.

\bibitem{36}    P.V. Moniz and J. Mour\~ao, Class.  Quantum Grav. {\bf 8}, (1991) 1815;
 
\bibitem{37}      J.M. Mour\~ao,  P.V. Moniz and 
P.M. S\'a, Class.  Quantum Grav. {\bf 10} (1993) 517;


G. Gibbons and  A. Steif, Phys. Lett. { \bf B320} 245 (1994); 


M.C. Bento, O. Bertolami, J.M. Mour\~ao,  P.V. Moniz and 
P.M. S\'a, Class.  Quantum Grav. {\bf 10} (1993) 285.


\bibitem{38}    O. Bertolami, Preprint Lisbon IFM-14/90, talk presented at the XIII International Colloquium on Group Theoretical Methods in Physics, Moscow, USSR June 1990, (Springer Verlag).

\bibitem{39} O. Bertolami, J.M. Mour\~ao, R.F. Picken and I.P. Volobujev, unpublished; 
S. Shabanov, talk presented at the First Iberian Meeting on Gravity, \'Evora, 
Portugal 
September 1992, edited by M.C. Bento, O. Bertolami, J.M. Mour\~ao and 
R.F. Picken (World Scientific Press, 1993); 


see also N. Manton, Ann. Phys. {\bf 167} (1986) 328; N. Manton, 
Nuc. Phys. {\bf B193} (1981) 502.



\bibitem{40} D.Z. Freedman and J. Schwarz, Phys. Rev. {\bf D15} (1977) 1007; 

S. Ferrara, F. Gliozzi, J. Scherk and P.v. Nieuwenhuizen, Nuc. Phys. {\bf 117} (1976) 333. 

\bibitem{nilp} See e.g., P.C. Aichelburg and R. G\"uven, Phys. Rev. Lett. {\bf 51} 
(1983) 1613;

 P. G.O. Freund, {\it Introduction to Supersymmetry}, 
(Cambridge U.P. -- 1986); 

B.S DeWitt, {\it Supermanifolds}, (Cambridge U.P.-- 1984); 

M. Henneaux and C. Teitelboim, {\it Quantization of Gauge Systems}, 
(Princeton U.P. -- 1992).

\bibitem{41}  C. Isham and J. Nelson, Phys. Rev. {\bf D10} (1974) 3226.

\bibitem{42}   T. Christodoulakis and J. Zanelli, Phys. Lett. {\bf 102A} (1984) 227; 

T. Christodoulakis and J. Zanelli, Phys. Rev. {\bf D29} (1984) 2738; 

T. Christodoulakis and C. Papadopoulos, Phys. Rev. {\bf D38} (1988) 1063

\bibitem{43}   P. D'Eath and J.J. Halliwell, Phys.~Rev.~D {\bf 35} (1987) 1100.


\bibitem{newnew} P. Moniz, {\em Rerum Universitas Sententia ex Susy}, 
essay-DAMTP R96/13;

{\em Conserved currents in supersymmetric quantum cosmology?}, 
DAMTP R96/14.

 \end{thebibliography}
\end{document}